\documentclass[11pt,onecolumn,floatfix,superscriptaddress]{revtex4}

\usepackage{graphicx}                   
\usepackage{epsfig}
\usepackage{amsmath,amssymb}            
\usepackage{hyperref}                   
\usepackage{epstopdf}
\usepackage{placeins}
\usepackage{upgreek}
\usepackage{slashed}
\usepackage{color}
\definecolor{orange}{rgb}{1,0.5,0}
\definecolor{brown}{rgb}{0.65, 0.16, 0.16}
\definecolor{phlox}{rgb}{0.87, 0.0, 1.0}

\graphicspath{{figs/}}                  

\begin{document}

    \title{CMB Polarization by the Asymmetric Template of Scalar Perturbations }
\author{Jafar Khodagholizadeh}
\email{gholizadeh@ipm.ir}
\affiliation{Farhangian University, P.O. Box 11876-13311, Tehran, Iran}
\author{ Rohoollah Mohammadi}
\email{rmohammadi@ipm.ir}
\affiliation{Iranian National Museum of Science and Technology (INMOST), PO BOX: 11369-14611, Tehran, Iran,\\
	School of Astronomy, Institute for Research in Fundamental Sciences (IPM), P. O. Box 19395-5531, Tehran, Iran.}
\author{ S. M. S. Movahed}
\email{m.s.movahed@ipm.ir}
\affiliation{Department of Physics, Shahid Beheshti University, 1983969411, Tehran, Iran.}

    \begin{abstract}
     Inspired by a dipole asymmetric template for the CMB temperature map in the primordial scalar fluctuations observed by Planck at a large scale, we examine the contribution of a similar template for power asymmetry in modifying the linear polarization pattern of CMB. Replacing un-modulated temperature fluctuation with dipolar modulated
one in time evolution equations somehow breaks linear perturbation in the various components of the CMB map. This non-linearity allows deflecting CMB polarization in patterns that contain divergence-free components. The explicit expressions for the angular power spectra of the electric and magnetic-type parities of linear polarization are derived in the form of the line of sight integral solutions.  Our results demonstrate that the electric-type polarization is modified and the magnetic-type polarization would be produced. Such imprints depend on the linear and square of the asymmetric amplitude for $E$- and $B$-modes power spectra, respectively. For the observed dipole template, the value of $B$ polarization spectrum at the large scale ($\ell\lesssim 10$) is almost equivalent to the power spectrum obtained from Compton scattering in the presence of tensor perturbation with tensor to scalar ratio about $r\simeq0.005$.
    \end{abstract}


    \maketitle
\section{Introduction}
  In the standard cosmology models, the homogeneity and isotropy of the universe are almost well-quantified assumptions and they essentially lead to CMB fluctuations behaving as the isotropic random field when the secondary anisotropies are well crossed out.
  Mentioned assumptions cause  many noticeable impacts not only on the computational methods pipelines but also lead to ruling out scenarios that give rise to anisotropic primordial fluctuations \cite{linde1997non,akrami2020planckinflation,dodelson2020modern,lesg13,Lesgourgues:2013qba,Ade:2015xua}. Due to the vital influences of homogeneity and isotropy, there are many researches have been focused on the examining mentioned property on  small and large scales fluctuations of the CMB map in both intensity and polarizations
   \cite[and references therein]{hajian2003measuring,hajian2006testing,copi2004multipole,copi2006large,hanson2009estimators,rath2013testing,rath2013direction,akrami2014power,planck2013results,Ade:2015hxq,akrami2020planck,rath2018testing,adhikari2018statistical,akrami2020plancknong,perivolaropoulos2021challenges}.
 Although, the previous extensive analysis indicated that the CMB field is consistent with the Gaussian prediction of $\Lambda$CDM scenario with statistical isotropy behavior \cite{akrami2020planck}, improving the accuracy and precision in the observing CMB stochastic field opened new rooms for the footprints of anomalies such as power asymmetry and deviation from statistical isotropy in a range of multipoles \cite[and references therein]{rath2013testing,rath2013direction,akrami2014power,planck2013results,Ade:2015hxq,rath2018testing,akrami2020planck,eriksen2004asymmetries,eriksen2005n,prunet2005constraints, hansen2009power, hoftuft2009increasing, Aluri:2017cna, Aiola:2015rqa}. Among the anomalies,  the hemispherical asymmetry parameterized by a dipolar modulation \cite{prunet2005constraints, gordon2005spontaneous} has been substantially investigated by different methods \cite{planck2013results,WMAP:2012fli,WMAP:2012nax}. A well-known template in the position space for dipolar modulation in the CMB temperature for the large scale reads as:
 \begin{equation}\label{eq:template}
    \tilde{\Delta}_T (\hat{n})=\Delta_T(\hat{n})[1+A_T \hat{P}.\hat{n}]
 \end{equation}
where $\Delta_T(\hat{n})\equiv \delta T(\hat{n})/\langle T\rangle$ is un-modulated temperature fluctuations in an arbitrary direction on the sky, $\hat{n}$, and $\hat{P}$ is the direction of dipolar modulation. The best-fit values for the amplitude of modulation in the temperature dipole and the corresponding preferred direction for low multi-poles $ \ell\in [2-64]$ have been reported as $A_T=0.072^{+0.031}_{-0.015}$ and $\hat{P}=(218 , -19)\pm29$  and for $ \ell\in [2-220]$, the amplitude is $A_T=0.023^{+0.008}_{-0.004}$ in direction $\hat{P}=(220 , -5)\pm25$ for joint analysis of $TT,TE,EE$ and for \texttt{Commander} observed data, respectively \cite{akrami2020planck}. In addition, the power asymmetry in the CMB polarization map has been examined in \cite{perivolaropoulos2021challenges,Aluri:2017cna,Namjoo:2014pqa,Ghosh:2018apx,Contreras:2017zjv}.
   Many phenomenological models have been proposed to elucidate observed asymmetry as of corresponding significance at the $\sim 3\sigma$ level such as the asymmetry in the initial condition of perturbations with different scenarios \cite{Namjoo:2013fka,Mazumdar:2013yta,mcdonald2014hemispherical,Byrnes:2016zxb,ashoorioon2016hemispherical,Agullo:2020fbw,Agullo:2020cvg,Marcos-Caballero:2019jqj}, super-horizon perturbations \cite{gordon2005spontaneous,gordon2007broken,Langlois:2008vk,Assadullahi:2014pya}, dipolar asymmetry due to cosmic string \cite{Jazayeri:2017szw}. The primordial  dipole asymmetries generated from different types of models could be  a source of CMB polarizations \cite{Namjoo:2014pqa}.

    In the standard scenario of cosmology, the contribution of primordial scalar perturbations in the generation of temperature anisotropies and linear polarization ($E$-mode) is dominant compared to other kinds of perturbations, while the $B$-mode polarization can not be generated by Compton scattering without considering the gravitational wave perturbations or non-linear density perturbation \cite{Mollerach:2003nq,Bartolo:2007dh, Fidler:2014oda}. 
    However, if we consider any interaction terms which can violate the Lorentz symmetry, it could be a nontrivial source of  $B$-mode polarization. More precisely, even with only asymmetric scalar perturbations, we expect to have an additional term with respect to the symmetric case and it has a footprint in the $B$-mode polarization. As an illustration, taking into account the photon-neutrino forward scattering without the tensor perturbation yields the $B$-mode polarization which can be significant for $50< \ell<200$ \cite{Mohammadi:2013dea,Khodagholizadeh:2014nfa}. The  Majorana dark matter can generate the $B$-mode in the presence of primordial scalar perturbations \cite{Mahmoudi:2018lll}   and it also has a contribution in cosmic  birefringence \cite{PhysRevD.108.023023}. Non-commutative space time framework can also generate the magnetic type polarization for CMB radiation \cite{Tizchang:2016vef}. The polarized Compton scattering is also a feasible approach to make a magnetic-like pattern in the linear polarization of CMB radiation \cite{ Khodagholizadeh:2019het}. In the presence of a homogeneous magnetic field, the Faraday rotation produces the CMB $B$-mode \cite{Scoccola:2004ke,Campanelli:2004pm,Kosowsky:2004zh}.  Also, if we have any conditions which can violate the Lorentz symmetry in the matter or the radiation distributions (like tensor perturbation of the matter or non-linear perturbation of radiation and matter), such conditions can generate $B$-mode linear polarization even by considering Thomson scattering \cite{Mollerach:2003nq,Bartolo:2007dh, Fidler:2014oda}.  Accordingly, by replacing $\Delta_T(\hat{n})$ which is un-modulated temperature fluctuation by dipolar modulated
    one $\tilde{\Delta}_T(\hat{n})$ in the time evolution equations, we somehow break the linear perturbations in the radiation which plays roles similar to the non-linear perturbations in radiation. The evolution of cosmological perturbations which breaks linear regime such as our considered dipole asymmetry template is characterized by mode-mixing, consequently it implies not only the different Fourier modes of temperature perturbations are influenced by dipole mode, but also the non-linear perturbations in CMB intensity (in the presence of Thompson scattering) can play as a source to deflect CMB linear polarization in the patterns containing divergence-free components. \\
    Motivated by detecting the asymmetry anomaly in the CMB data modeled by a viable template, Eq. (\ref{eq:template}), and various sources of magnetic-type CMB polarization, we would like to examine the influence of taking into account a similar template for initial scalar perturbations in the Fourier space on the CMB polarization. We will show that the imprint of such an asymmetric model on the magnetic-type CMB polarization for small $\ell$ is almost equivalent to the $B$-mode produced by primordial tenor perturbations and it is proportional to the square value of asymmetry amplitude.
    \\
    The rest part of this paper is organized as follows: considering the dipole asymmetry in the initial perturbations, we revisit the Boltzmann equations of CMB temperature and polarization anisotropies in an exact approach in section \ref{Sec:method}. Section \ref{sec:3} is devoted to calculating the $E$-mode and $B$-mode power spectra. We also compare our results with that achieved by considering the primordial tensor perturbation. Summary and conclusions will be given in section \ref{sec:4}. To make our analysis more sense, we will derive some important equations in Appendices.
   \section{Methodology}\label{Sec:method}
\begin{figure}
    \begin{center}
        \includegraphics[scale=0.78]{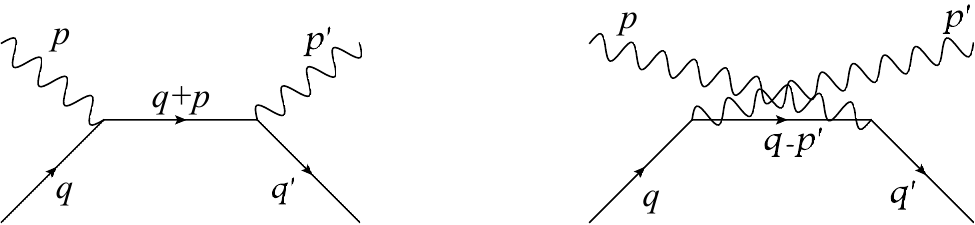}
    \end{center}
    \caption{Full detail of the Compton scattering Feynman diagrams in the presence of $H_I^0(t)$.} \label{fig2}
   \end{figure}

    In the standard scenario, the symmetric scalar perturbation at the linear order can not generate the $B$-mode \cite{Kosowsky:1994cy,Zaldarriaga:1996xe}. While incorporating the asymmetric part for scalar perturbation may produce nontrivial terms in the evolution of Stokes parameters which effectively indicates the non-linear order of interactions. In this section, we will rely on a famous asymmetry template for the scalar fluctuations, to examine the generating CMB polarization.
    We start with the scattering theory and try to calculate the time evolution of the quantum number density of the photons including local interaction as perturbation term in the associated Hamiltonian according to $H=H_0+H_I$.  Now, we pursue the second-quantized formalism with creation and annihilation operators for the photons and electrons obeying the canonical commutation relations as:
\begin{eqnarray}\label{commutator122}
[a_{s}(p),a_{s^{'}}^{\dagger}(p^{'})]=(2\pi)^{3}2 p^{0}\delta^{3}(\bold{p}-\bold{p^{'}})\delta_{ss^{'}}\\
\lbrace b_{r}(p),b_{r^{'}}^{\dagger}(p^{'})\rbrace=(2\pi)^{3}\dfrac{q^{0}}{m} \delta^{3}(\bold{q}-\bold{q^{'}})\delta_{r r^{'}}\nonumber
\end{eqnarray}
where $s$ and $s'$ labels denote the photon polarization while the $r$ and $r'$ labels refer to the electron spin. The bold momentum variables represent three-momenta while plain momentum variables represent
four-momenta.
The photon density matrix incorporating the linear and circular polarizations reads as \cite{Kosowsky:1994cy}:
        \begin{eqnarray}\label{eq:rho}
        \hat{\rho}=\int \dfrac{d^{3}k}{(2\pi)^{3}}\rho_{ij}(k)D_{ij}(\vec{k})
        \end{eqnarray}
        here $ D_{ij} (\vec{k})\equiv a_{i}^{\dagger}(\vec{k})  a_{j}(\vec{k}) $ is the photon number operator written in the Fourier space and $ \rho_{ij} $ is the density matrix.
        The expectation value of $D$ is proportional to density matrix. Direct calculation shows:
        \begin{eqnarray}
        \langle D_{ij}\rangle=tr[\hat{\rho}D_{ij}]=\int \frac{d^3p}{(2\pi)^3}\langle p|\hat{\rho}D_{ij}(\bold{k})|p\rangle=(2\pi)^3\delta^{(3)}(0)2k^0\rho_{ij}(\bold{k})
        \end{eqnarray}
        The right-hand side of the above equation achieves from the successive implementation of the commutation relation Eq.(\ref{commutator122}); the infinite delta function results from the infinite quantization volume necessary with continuous momentum variables, and cancels out all
        physical results.
        According to the Stokes parameters, the photon density matrix also becomes:
        \begin{eqnarray}\label{eq:rho1}
        \rho_{ij}\equiv\frac{1}{2}\left(
        \begin{matrix}
        T+Q & U-iV \\
        U+iV&  T-Q \\
        \end{matrix}
        \right)
        \end{eqnarray}
        The number density operator evolution equation in the presence of perturbation term in Hamiltonian and considering the free fields assumption   is given by the quantum Boltzmann equation as \cite{Kosowsky:1994cy}
        \begin{eqnarray}\label{evolution1}
        (2\pi)^3\delta^3(0)2k^0\dfrac{d}{dt}\rho_{ij}(\vec{k})=i\langle[H_{I}^0(t);D_{ij}^{0}(\vec{k})]\rangle -\dfrac{1}{2}\int_{-\infty}^{+\infty} dt \langle[H_{I}^0(t);[H_{I}^0(t);D_{ij}^{0}(\vec{k})]]\rangle
        \end{eqnarray}
        Now, we can calculate the first order of interaction Hamiltonian of QED ($H_{I}^0(t)$), as a function of the free fields for photon-electron scattering as (see Fig. (\ref{fig2})):
     \begin{eqnarray}
H_I^0(t)&=&\int d\bold{q}d\bold{q'}d\bold{p}d\bold{p'}(2\pi)^3\delta^3(\bold{q'+p'-q-p})\exp[it(q'^0+p'^0-q^0-p^0)]\\
&\times&[b^\dagger_{r'}(q')a^\dagger_{s'}(p')(\mathcal(M_1+M_2))a_s(p)b_r(q)],\nonumber
    \end{eqnarray}
  Where $\bold{p} $ and $ \bold{q} $ are incoming photon momentum vector  and electron momentum vector respectively, $ \vert \bold{p} \vert = p^0   $ and   $ \vert \bold{q} \vert = q^0   $. It is noticed that the prime of these will be the outgoing photons and electrons.  Also, $ M_1  $ and $ M_2 $ are scattering amplitudes which are shown as below
    \begin{eqnarray}
    M_1(q'r',p's',qr,ps)\equiv e^2\frac{\bar{u}_{r'}(q')\epsilon\!\!\!/_{s'}(p')(p\!\!\!/+q\!\!\!/+m)\epsilon\!\!\!/_{s}(p)u_r(q)}{2p.q }\nonumber\\
    M_2(q'r',p's',qr,ps)\equiv e^2\frac{\bar{u}_{r'}(q')\epsilon\!\!\!/_{s}(p)(q\!\!\!/-p'\!\!\!/+m)\epsilon\!\!\!/_{s'}(p')u_r(q)}{2p'.q }
    \end{eqnarray}
with the abbreviations
 \begin{eqnarray}
     d\bold{q}\equiv\frac{d^3\bold{q}}{(2\pi)^3}\dfrac{m}{q^0}~~,~~ d\bold{p}\equiv\frac{d^3\bold{p}}{(2\pi)^3p^0}
 \end{eqnarray}
for  electrons and photons respectively. Where $ u_r $ is a spinor solution to the Dirac equation with spin index $ r=1,2 $ and $ \epsilon_{s} $ are photon polarization four-vector with index $ s=1,2$. Also  $r, r'$ are incoming and outgoing electron spin indices and $ s, s'$ are  the exactly same for photons.
    In addition, the first and second terms on the right-hand side of Eq. (\ref{evolution1}) are the forward scattering and the higher-order collision terms, respectively.
     We proceed with our calculation by emphasizing the polarization sums explicitly and up to the first order in scattering terms. After tedious but straightforward calculation, we have \cite{Kosowsky:1994cy}:
        \begin{eqnarray}\label{eq:apprhoij}
        \dfrac{d}{dt}\rho_{ij}(\vec{x},\vec{k})&=&\dfrac{e^{4}n_{e}(\vec{x})}{16\pi m^{2}k}\int_{0}^{\infty}p dp \int \dfrac{d\Omega_p}{4\pi}\left[\delta (k-p)+(\vec{k}-\vec{p}). \vec{v}(\vec{x})\dfrac{\partial \delta (k-p)}{\partial p} \right] \nonumber\\ &\times & \big[ -2\left (\dfrac{p}{k}+\dfrac{k}{p}\right )\rho_{ij}(\vec{x},\vec{k})+ 4\hat{p}.\hat{\varepsilon}_{i}(\vec{k}) \hat{p}.\hat{\varepsilon}_{1}\rho_{1j}(\vec{x},\vec{k})+4\hat{p}.\hat{\varepsilon}_{i}(\vec{k}) \hat{p}.\hat{\varepsilon}_{2}\rho_{2j}(\vec{x},\vec{k})\nonumber\\
        &+&\left(\dfrac{p}{k}+\dfrac{k}{p}-2\right)\delta_{ij}(\rho_{11}(\vec{x},\vec{p})-\rho_{22}(\vec{x},\vec{p}))\nonumber\\
        &+&\left(\dfrac{p}{k}+\dfrac{k}{p}\right)\left(\varepsilon_{i}(\vec{k}).\varepsilon_{1}(\vec{p})\varepsilon_{j}(\vec{k}).\varepsilon_{2}(\vec{p})-\varepsilon_{i}(\vec{k}).\varepsilon_{2}(\vec{p})\varepsilon_{j}(\vec{k}).\varepsilon_{1}(\vec{p})\right)(\rho_{12}(\vec{x},\vec{p})-\rho_{21}(\vec{x},\vec{p}))\nonumber\\&+& 2\left(\varepsilon_{i}(\vec{k}).\varepsilon_{1}(\vec{p})\varepsilon_{j}(\vec{k}).\varepsilon_{2}(\vec{p})+\varepsilon_{i}(\vec{k}).\varepsilon_{2}(\vec{p})\varepsilon_{j}(\vec{k}).\varepsilon_{1}(\vec{p})\right)(\rho_{12}(\vec{x},\vec{p})+\rho_{21}(\vec{x},\vec{p}))\nonumber\\&+& 4 \varepsilon_{i}(\vec{k}).\varepsilon_{1}(\vec{p})\varepsilon_{j}(\vec{k}).\varepsilon_{1}(\vec{p})\rho_{11}(\vec{x},\vec{p})+4 \varepsilon_{i}(\vec{k}).\varepsilon_{2}(\vec{p})\varepsilon_{j}(\vec{k}).\varepsilon_{2}(\vec{p})\rho_{22}(\vec{x},\vec{p})\big],
        \end{eqnarray}
         here $\vec{v}(\vec{x})$ is the electron bulk velocity. Using  Eqs. (\ref{eq:rho}), (\ref{eq:rho1}) and (\ref{eq:apprhoij}), the collisional term of the Boltzmann equation for the polarization parts can be written as:
         \begin{eqnarray}\label{eq:appcollision}
         \dot{\Delta}_{Q}&=&\sigma_{T}\int\dfrac{d\Omega_{p}}{4\pi}[F_{QT}^{p}(\Omega_{p})\Delta_{T}(\vec{p}) + F_{QT}^{k}(\Omega_{p})\Delta_{T}(\vec{k})+F_{QU}^{p}(\Omega_{p})\Delta_{U}(\vec{p})+F_{QQ}^{p}(\Omega_{p})\Delta_{U}(\vec{p})],\label{qdot}\nonumber\\
         \dot{\Delta}_{U}&=&\sigma_{T}\int\dfrac{d\Omega_{p}}{4\pi}[F_{UT}^{p}(\Omega_{p})\Delta_{T}(\vec{p}) + F_{UT}^{k}(\Omega_{p})\Delta_{T}(\vec{k})+F_{UU}^{p}(\Omega_{p})\Delta_{U}(\vec{p})+F_{UQ}^{p}(\Omega_{p})\Delta_{U}(\vec{p})],\label{udot}
         \end{eqnarray}
         where $\dot{\Delta}_{Q}\equiv\frac{d}{d\tau}\,\Delta_{Q}$ and $\dot{\Delta}_{U}\equiv\frac{d}{d\tau}\,\Delta_{U}$. To avoid the messy mathematical formula, we move the detailed explanation of variables in Appendix A.
         Now, we rely on a famous asymmetric template for the scalar fluctuations, to examine the CMB polarizations. To this end, we start with  the asymmetric template for scalar perturbations in the Fourier space which is used in the hierarchical Boltzmann equation as:
   \begin{equation}\label{eq:tamplate}
    \tilde{\Delta}_{T} (\vec{k},K,\tau)= \Delta_{ T}(\vec{k},K,\tau)(1+A_T\,\hat{P}.\hat{k})
   \end{equation}
   where the $\Delta_{T}(\vec{k},K,\tau)$ is fluctuation contrast at direction $\hat{n}=\vec{k}/k$ to the line of sight and $\hat{P}$ is the assumed direction of the dipole asymmetry. Also, $\tau$ and $K$ are the conformal time and the value of comoving Fourier mode, respectively. To take into account the contribution of dipole asymmetry (Eq. (\ref{eq:tamplate})), the $\Delta_{T} (\vec{p},K)$ and $\Delta_{T} (\vec{k},K)$ in Eq. (\ref{eq:appcollision}) should be replaced by $ \tilde{\Delta}_{T} (\vec{p},K)= \Delta_T(\vec{p},K)(1+A_T\hat{P}\cdot\hat{p}) $ and $ \tilde{\Delta}_{T} (\vec{k},K)= \Delta_T(\vec{k},K)(1+A_T\hat{P}.\hat{k})$,  respectively. Therefore, the evolution equation for the Stokes parameters, $Q^{(S)}$ and $U^{(S)}$ would be modified in the presence of dipole asymmetric scalar perturbation (see Appendix A for more details):
    \begin{eqnarray}
    \frac{d}{d\tau}(\tilde{Q}^{(\rm S)}\pm i\tilde{U}^{(\rm S)}) +iK\mu (\tilde{Q}^{(\rm S)}\pm i\tilde{U}^{(\rm S)}) &=& -\dot\kappa\Large[
    (\tilde{Q}^{(\rm S)}\pm i\tilde{U}^{(\rm S)}) +{1\over 2} [1-P_2(\mu)] \Pi-\,\Pi^{\pm (\rm S)}\Large];
    \label{eq:Boltzmann1}
    \end{eqnarray}
   where  $\Pi\equiv \Delta_{T2}^{(\rm S)} +\Delta_{P2}^{(\rm S)}+\Delta_{P0}^{(\rm S)}$ and $(S)$ denotes the scalar perturbations of matter. The differential optical depth for Thomson scattering is denoted by $\dot{\kappa}=an_ex_e\sigma_T$, where $a(\tau)$ is the scale factor normalized
   to unity at the present as a function of conformal time ($\tau$). The electron density and the ionization
   fraction are denoted by $n_e$ and $x_e$, respectively.  Also, $\sigma_T$ is the Thomson cross-section.
   The above equation without considering the last term on the right side, e.g $ \Pi^{\pm (\rm S)}(k,K,\tau)$, is a general Boltzmann equation for the linear polarization for CMB,
   which can not generate $B$-mode as carried out in the context of standard scenario \cite{Kosowsky:1994cy, Zaldarriaga:1996xe,Bond:1984fp}. Note replacing $ \tilde{\Delta}_{T} (\vec{k},K,\tau)$ in the Boltzmann equation somehow breaks the linear regime of radiation perturbation and  this non-linearity in CMB temperature fluctuations generates the nontrivial terms (last terms in the right-hand side of Eq. (\ref{eq:Boltzmann1})) which can deflect CMB linear polarization in the patterns containing divergence-free
   components. The mentioned nontrivial terms are given by:
   \begin{equation}\label{eq:Apm}
    \Pi^{\pm (\rm S)}(k,K,\tau)=A_T\int \frac{d\Omega_p}{4\pi}\,(\hat{p}\cdot\hat{P})\,\Delta_T(\vec{p},K,\tau)\,(F_{QT}^p\pm i\,F_{UT}^p),
   \end{equation}
   also by using $\hat{p}\cdot\hat{P}=\frac{4\pi}{3}\sum\,Y^*_{1,m}(\hat P)Y_{1,m}(\hat p)$, the Eq. (\ref{eq:Apm}) reads as:
   \begin{equation}\label{Apm1}
    \Pi^{\pm (\rm S)}(k,K,\tau)=A_T\,\sum_{m=-3}^3\,\Delta_{T,3m}(k,K,\tau)\,F^\pm_{3m},
   \end{equation}
   the $F_{QT}^p$, $F_{UT}^p$ and $ F^\pm_{3m}$ are respectively defined  by Eqs. (\ref{eq:appF}) and Eqs. (\ref{eq:appf3}) presented in Appendix A. Also $\Delta_{3m,T}(k,K,\tau)\equiv\int d\Omega_k  Y^*_{3m}(\hat{k}) \,\Delta_T (\vec{k},K,\tau)$. The $\hat{p}$ and $\hat{n}$ are the directions of photon propagation and the line of sight, respectively.
   Considering the $\Delta_P^{\pm(\rm S)}=Q^{\pm(\rm S)}\pm iU^{\pm(\rm S)}$ for polarization anisotropy in the context of dipole asymmetry template for the scalar perturbations, we can separate $\Delta_P^{\pm(\rm S)}$ into the symmetric and asymmetric parts as:
   \begin{eqnarray}\label{eq:DA00}
    \tilde{\Delta} _{P}^{\pm (\rm S)}(\hat{{n}})
    &=&\Delta _{P}^{\pm (\rm S)}(\hat{{n}})\,+\,\Delta _{P}^{\pm (\rm S)}(\hat{{n}})\Big|_{\rm Asymmetry},\label{DA1}
   \end{eqnarray}
   \begin{equation}\label{eq:DA01}
    \Delta _{P}^{\pm (\rm S)}(\hat{{n}})\Big|_{\rm Asymmetry}=\int d^3 \vec{K} \xi(\vec{K})e^{\mp 2i\phi_{K,n}} \Delta _{P}^{\pm (\rm S)}(K,\mu,\tau_0)\Big|_{\rm Asymmetry}
   \end{equation}
   where
   \begin{eqnarray}
    \Delta _{P}^{\pm (\rm S)}(K,\mu,\tau_0)\Big|_{\rm Asymmetry}
    &=&\int_0^{\tau_0} d\tau\, g(\tau)\,
    e^{ix \mu }\,\,\Pi^{\pm (\rm S)}(K,\tau).\label{DA2}
   \end{eqnarray}
   where $\mu\equiv |\hat{K}.\hat{k}|$, $g(\tau)=\dot{\kappa} \exp({\kappa})$ is visibility function which is written in terms of optical depth, $\kappa$. The differential optical depth is  The differential optical depth for Thomson scattering is also denoted by $\dot{\kappa}=an_ex_e\sigma_T$ and $x= K (\tau_0-\tau)$. The $\vec{K}$ and $\hat{{n}}$ can be rotated to a fixed frame in the sky by the angle named by $\phi_{K,n}$ and $\xi(\vec{K})$
   is a random variable used to characterize the initial
   amplitude of the $\vec{k}$-mode which has the following statistical property
   as $
   \langle \xi^{*}({\vec{K}}_1)\xi({\vec{K}}_2)
   \rangle=
   \mathcal{P}_{\xi}({\vec{K} })\delta_{D}({\vec{K}}_1- {\vec{K}}_2),
   $
   where $\mathcal{P}_{\xi}(K)$ is initial power spectrum which depends only on the magnitude $K$ of the wave vector $\vec{K} $. The $\delta_{D}$ is Dirac delta function.
   \section{Dipole asymmetry impact on the E- and B-modes}\label{sec:3}
   In this section, by using our methodology presented in the previous section, we turn to compute the $E$- and $B$-modes power spectra modified by a dipole asymmetric template for scalar perturbations. One can expand $\tilde{\Delta} _{P}^{\pm (\rm S)}(\hat{n})$ (Eq. (\ref{eq:DA00})) in the appropriate spin-weighted basis, $\tilde{a}_{\diamond,\ell m}$, resulting in:
   \begin{eqnarray}
    \tilde{a}_{E,\ell m}=\left[\frac{(\ell+2)!}{(\ell-2)!}\right]^{-\frac{1}{2}}\int d\Omega Y_{\ell m}^{\ast}[\bar{\eth}^{2}\tilde{\Delta} _{P}^{+(S)}(\hat{n})+\eth^{2}\tilde{\Delta} _{P}^{- (S)}(\hat{n})],\nonumber\\
    \tilde{a}_{B,\ell m}=\left[\frac{(\ell+2)!}{(\ell-2)!}\right]^{-\frac{1}{2}}\int d\Omega Y_{\ell m}^{\ast}[\bar{\eth}^{2} \tilde{\Delta} _{P}^{+ (S)}(\hat{n})-\eth^{2}\tilde{\Delta} _{P}^{- (S)}(\hat{n})], \label{e-b modes2}
   \end{eqnarray}
   separating to the symmetric and asymmetric parts leads to $\tilde{a}_{\diamond,\ell m}=a_{\diamond,\ell m}+a_{\diamond,\ell m}\Big|_{\rm Asymmetry}$ where the $\diamond$ can be replaced by $E$- and $B$-modes. The asymmetric part reads as:
   \begin{eqnarray}
    a_{\diamond,\ell m}\Big|_{\rm Asymmetry}=\left[\frac{(\ell+2)!}{(\ell-2)!}\right]^{-\frac{1}{2}}\int d\Omega Y_{\ell m}^{\ast}\left[\bar{\eth}^{2}\Delta _{P}^{+(\rm S)}(\hat{n})\Big|_{\rm Asymmetry}\pm\eth^{2}\Delta _{P}^{- (\rm S)}(\hat{n})\Big|_{\rm Asymmetry}\right], \label{e-b modesA1}
   \end{eqnarray}
   the ``$\pm$" is replaced by ``$+$" for $E$-mode and  by ``$-$" for $B$-mode. The associated power spectra are given by $\tilde{C}_{\diamond\diamond,\ell}^{(\rm S)}=\frac{1}{2\ell+1}\sum_{m=-\ell}^{\ell} \langle\tilde{a}^*_{\diamond,\ell m} \,\tilde{a}_{\diamond, \ell m}\rangle$. Finally the leading order terms of the $E$- and $B$-modes power spectra in the presence of scalar perturbations with a dipole asymmetry are:
   \begin{eqnarray}\label{Emode1}
    &&C_{EE,\ell}^{(\rm S)}\Big|_{\rm Asymmetry}= \dfrac{8\pi}{2\ell+1} \frac{(\ell-2)!}{(\ell+2)!}\int K^{2}dK \mathcal{P}_{\xi}(K) \nonumber \\
    &&~~~~~~~~~~~~~~\times\sum_{m=-\ell}^{\ell}  \left[\int d\Omega Y_{\ell m}^{\ast}(\hat{n})\int_{0}^{\tau_{0}}d\tau \, g(\tau)\, [\bar{\eth}^{2}\Pi^{+(\rm S)}+\eth^{2}\Pi^{- (\rm S)}]e^{ix\mu}\right]^* \\&&~~~~~~~~~~~~~~~~~~~~~~~~~~~~~~~~~~~  \left[\int d\Omega Y_{\ell m}^{\ast}(\hat{n}) \int_{0}^{\tau_{0}}d\tau \, g(\tau)\,\Pi\,\partial_{\mu}^{2}[(1-\mu^{2})(1-P_2) e^{ix\mu}]\right]  \nonumber\\ \nonumber
    \label{EModeM}
   \end{eqnarray}
   and
\begin{eqnarray}\label{eq:Bmode1}
    C_{BB,\ell}^{(\rm S)}\Big|_{\rm Asymmetry}&=& \dfrac{(4\pi)}{2\ell+1} \frac{(\ell-2)!}{(\ell+2)!}\int K^{2}dK \mathcal{P}_{\xi}(K) \sum_m \Big|\int d\Omega Y_{\ell m}^{\ast}(\hat{n}) \int_{0}^{\tau_{0}}d\tau \, g(\tau)\, [\bar{\eth}^{2}\Pi^{+(\rm S)}-\eth^{2}\Pi^{- (\rm S)}]e^{ix\mu}\Big|^2\nonumber \\
   \end{eqnarray}
   here $g(\tau)=\dot{\kappa}e^{\kappa}$ is visibility function. For the symmetric scalar perturbations, there is no source to generate $B$-mode polarization  as we expect, while in our case, due to the asymmetric template for the scalar perturbations, the $B$-mode is generated even without any tensor perturbations. After doing some mathematical derivations, Eq. (\ref{eq:Bmode1}) becomes:
   \begin{eqnarray}\label{14}
    C_{BB,\ell}^{(\rm S)}\Big|_{\rm Asymmetry}&=& (4\pi)^{2} \frac{(\ell+2)!}{(\ell-2)!} \int K^{2}dK P_{\varphi}(K) \left[\int_{0}^{\tau_{0}}d\tau \, g(\tau)\, A(K,\tau)\Big|_{\rm Asymmetry} \, \dfrac{(\ell-2)j_{\ell}(x)-xj_{\ell+1}(x)}{x^{3}}\right]^2,\quad\quad \label{BModeM}
   \end{eqnarray}
   \begin{figure}
    \begin{center}
        \includegraphics[scale=0.75]{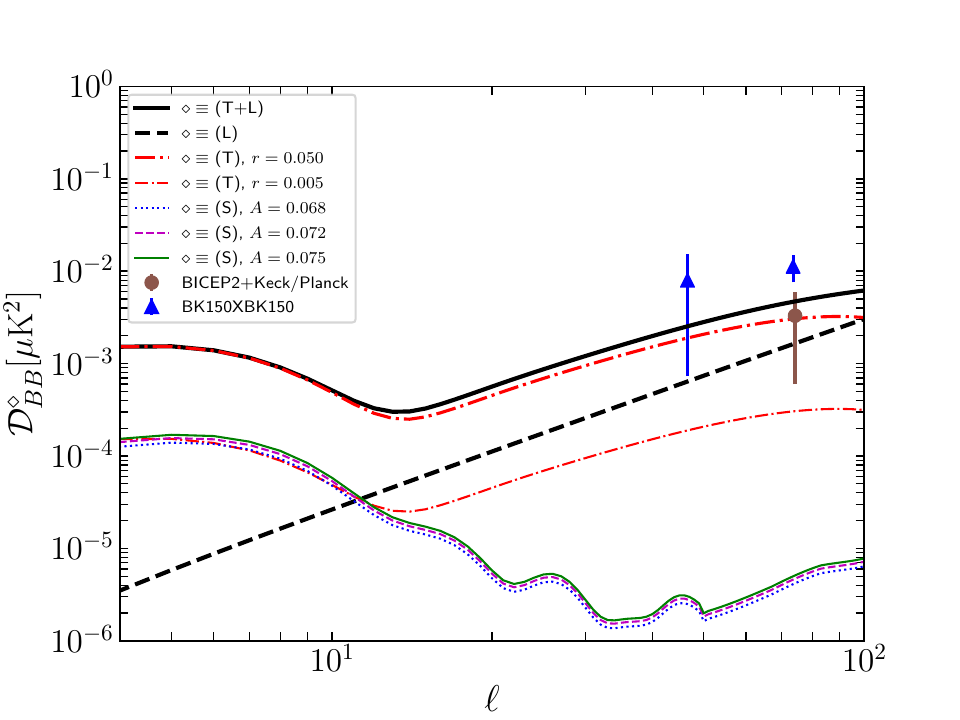}
        \includegraphics[scale=0.75]{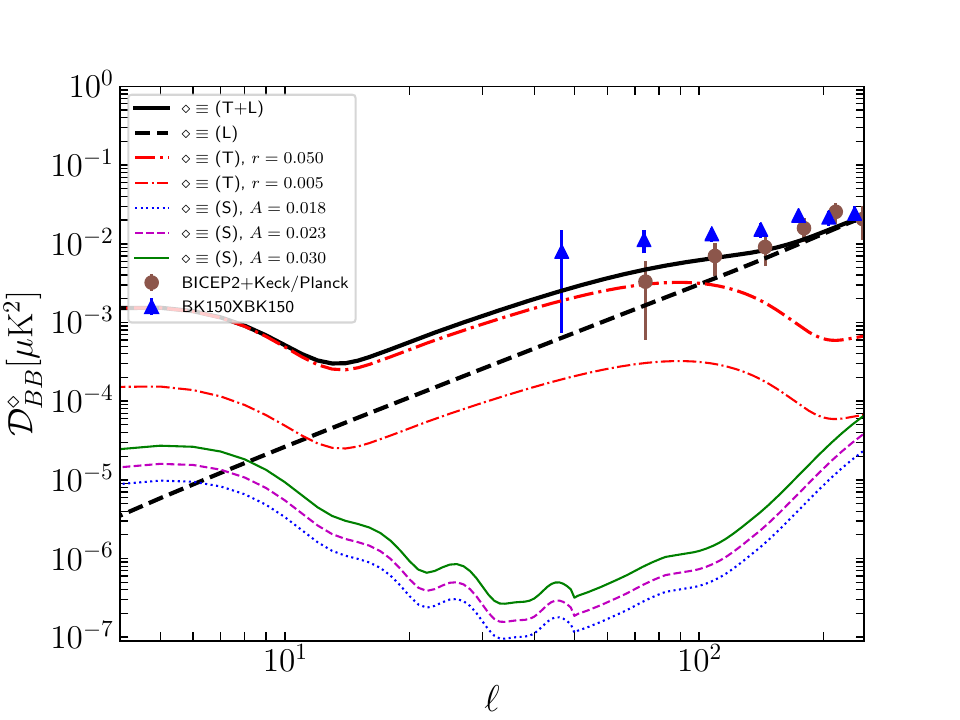}
    \end{center}
    \caption{The $B$-mode power spectra for different components. The $C^{(\rm T)}_{BB,\ell}$
    is for tensor mode (thick dash-dot line for $r=0.05$ and thin dashed-dot curve for
$r=0.005$. The thick dashed line indicates the contribution of
lensing ($C^{(\rm L)}_{BB,\ell}$)). The thick solid line corresponds
to a summation of tensor ($r=0.05$) and lensing parts ($C^{(\rm
T+L)}_{BB,\ell}$). The thins curves are devoted to the $B$-mode
generated by Compton scattering in the presence of dipole asymmetry
in the scaler fluctuations for various amplitudes (the upper panel is
for $A_T=0.068, 0.072\; \& \; 0.075$ when $\ell\in[2,100]$ while the
lower panel is  for $A_T=0.018, 0.023\; \& \;0.030$ when
$\ell\in[2,250]$). The filled circle and triangle symbols are
associated with BICEP2+Keck Array/Planck and BICEP2+Keck at 150 GHz
band, respectively.} \label{fig:fig1}
   \end{figure}
where  (see the Appendix B for more details):
\begin{eqnarray}\label{AA}
A(K,\tau)\Big|_{\rm Asymmetry} &=& 0.753\,  A_T\,\Delta_{T,3m}(k,K,\tau)
\end{eqnarray}
   The amplitude and the direction of the low-$\ell$ dipole asymmetry signal have been determined from quadratic maximum likelihood in the range $\ell\in[2,65]$  from joint analysis of $TT,TE,EE$ and for \texttt{Commander} observed by $Planck$. The corresponding values at $1\sigma$ confidence interval are $A_T=0.072^{+0.031}_{-0.015}$ and $\hat{P}=(218 , -19)\pm29$  \cite{akrami2020planck}. On the other hand, for the range $ \ell\in[2,220]$, the observational constraint on the amplitude is $A_T=0.023^{+0.008}_{-0.004}$ for the direction  $\hat{P}=(220, -5)\pm25$.
Our results demonstrate that additional contributions are assigned to the $E$- and $B$-modes power spectra sourced by $\Delta_{3,T}$. In addition, the power spectrum of $E$- and $B$-modes have a linear and square dependency on $A(K,\tau)\Big|_{\rm Asymmetry}$,  respectively. In another word, mentioned results illustrate a lower bound on the $B$-mode irrespective of the existence of primordial and secondary sources which generate the CMB $B$-mode. Comparing the  Eq. (\ref{14}) with the dominant part of standard linear polarization represented by $\bar{C}_{EE,\ell}^{(\rm S)}$, one can deduce an upper limit for the $B$-mode generated by asymmetric scalar perturbations such that $C_{BB,\ell}^{(\rm S)}\Big|_{\rm Asymmetry}\lesssim \left(  A(K,\tau)\Big|_{\rm Asymmetry} \bar{C}_{EE,\ell}^{(\rm S)}\right)$. \\
  In Fig. \ref{fig:fig1}, we compute the $B$-mode power spectra ($\mathcal{D}_{BB}\equiv \ell (\ell+1)C_{BB,\ell}/2\pi$) for the different components. The $C^{(\rm{T})}_{BB,\ell}$ and $C^{(\rm{L})}_{BB,\ell}$ correspond to the tensor (T) and lensed-$\Lambda$CDM (L) modes, respectively.  The $C^{(\rm{S})}_{BB,\ell}$ is for the $B$-mode power spectrum generated by the dipole asymmetry in the scale perturbations with different amplitudes (Eq. (\ref{BModeM})). The filled circle symbols are devoted to the dust-subtracted $B$-mode power provided by BICEP2+Keck data and $Planck$ Joint analysis \cite{BICEP2:2015nss}\footnote{\texttt{https://lambda.gsfc.nasa.gov}}, while the filled triangle symbols illustrate the BICEP2+Keck auto-correlation at 150 GHz band \cite{BICEP2:2018kqh} \footnote{\texttt{http://bicepkeck.org/}} .
   Interestingly, for small $\ell$, the value of $B$-mode spectrum generated by Compton scattering in the presence of dipole asymmetry in the scalar perturbations is almost equivalent to the $B$-mode due to the tensor perturbations for $r\simeq 0.005$. The asymmetric template has a dominant contribution for large angular scale (small $\ell$), therefore for $\ell\in[2,100]$ the observational constraint leads to a higher value for asymmetric amplitude compared to taking into account the higher multiples or small scales $\ell\in[2,250]$. To describe the various physical influences on the CMB fluctuations, in principle, the Boltzmann-Einstein equations governing the evolution of the anisotropies in the cosmic distribution of photons as well as matter inhomogeneities should be solved. The CMB anisotropies are sourced by the primary and secondary signatures. However, the mentioned framework provides a straightforward approach to tracing the various phenomena in generating CMB anisotropies. Still, from the observational point of view, it is helpful to rely on the probabilistic approach constructed for quantifying specific stochastic fields such as CMB. The CMB two-point correlation functions and corresponding expansions in Legendre polynomials and spherical harmonics denoted by power spectra are practically well-defined observables. The asymmetric part of $C_{BB,\ell}^{(\rm{S})}$ (Eq. (\ref{14})), reveals that the initial condition (power spectrum of initial fluctuations) is convolved by a transfer function determined by the Boltzmann equation modified due to the presence of asymmetric template for scalar perturbations. The modified transfer function includes the integral form of the generalized amplitude of the asymmetric template (Eq. (\ref{AA})) multiplied by the visibility function and a functional form of the spherical Bessel function. Comparing the mentioned part with the standard $E$- and $B$-modes power spectra indicates a significant discord in addition to the new functional form of spherical Bessel function, which is implying the presence of hexapole temperature fluctuations instead of quadrupole. The contribution of the asymmetric part for the almost large scale, $\ell\lesssim [10-20]$, looks like the $B$-mode power spectrum generated by tensor perturbation without any gravitational redshift term. At this scale, the contribution of Thomson scattering plays a curtail role in the polarization spectra. To carry out a more precise evaluation of the B-mode spectrum at intermediate and small scales due to the given asymmetric template for scalar perturbations, we should take into account the better approximation done for the $A(K,\tau)$ (Eq. (\ref{AA})) which is acceptable for determining the upper limit on the B-mode at large scale. However, the oscillation behavior of  B-mode is due to the spherical Bessel function, the visibility function accompanying the generalized amplitude of the asymmetric template of scalar perturbations having a sharp maximum at $\ell=K(\tau_0-\tau)$. 
    Mentioned behavior is illustrated in the upper and lower panels of Fig. \ref{fig:fig1}. The upper panel is for $A_T=0.068, 0.072\; \& \; 0.075$ when $\ell\in[2,100]$ while the lower panel is  for $A_T=0.018, 0.023\; \& \;0.030$ when $\ell\in[2,250]$.

   \section{ Summary and Conclusion}\label{sec:4}
   In this study, inspired by the observed asymmetric template for the CMB temperature power spectrum on large scales, we considered the asymmetric part for the scalar perturbations. We revisited the quantum Boltzmann equations for the density matrix of the CMB temperature as well as the CMB Stokes parameters via Compton scattering in the presence of a dipole asymmetric template for the scalar perturbations. In the Boltzmann equations, we have replaced un-modulated temperature fluctuations $\Delta_T(\hat{n})$ by dipolar modulated
   one $\tilde{\Delta}_T(\hat{n})$ which breaks the linear perturbations in the radiation and plays a role similar to the non-linear perturbations in radiation.
   We derived the explicit expressions for the angular power spectra of the electric and magnetic-types parities of linear polarization in the form of the line of sight integral solutions and finally, we obtained the departure from the results given by considering the symmetric template for the scalar perturbations.

   Our results demonstrated that scalar fluctuations with a dipole asymmetric template modified the standard Boltzmann equations for the linear polarization of the CMB map. The $E$-mode received an additional term with a linear dependency on the amplitude of the temperature asymmetric template according to Eqs. (\ref{Emode1}). We obtained the $B$-mode power spectrum achieved a contribution via Compton scattering in the presence of the primordial scalar fluctuations with dipole asymmetric template which not be zero in contrast with the contemporary model. As shown in Eq. (\ref{BModeM}), $C_{BB,\ell}^{(\rm S)}$ depends on the square of dipole asymmetry amplitude.
   To make more sense, we compared the generated $B$-mode due to asymmetric scalar perturbations with magnetic-type parity of linear polarization in the presence of tenor perturbations with tensor to scalar ratio about $r\simeq0.005$, which are almost equivalent for $\ell\lesssim 10 $ (Fig. \ref{fig:fig1}).\\
   It could be interesting to assess the contributions of  $\ell=3$ multi-poles, $\Delta_{T,3m}(k,K,\tau)$, more precisely, to achieve a more accurate estimation of the $B$-mode instead of our current result which is an upper estimation. Also applying the computational data modeling to put a precise constraint on the model-free parameters improves our results \cite{Ghosh:2018apx}. 

   \section*{Acknowledgment}
   This research is supported in part by INSF with Grant No: 95838970.   We are grateful to Sorhab Rahvar  for his constructive comments.

   \section*{Appendices}

   \subsection*{Appendix A}\label{appA}
    Accounting for the scalar mode perturbations and by using the Thomson scattering of CMB photons by cosmic electrons, the time evolution of $\rho_{ij}(\vec{x},\vec{k})$ (Eq. (\ref{eq:rho})) as well as the Stokes parameters (Eq. (\ref{eq:rho1})) is given by Eq. (\ref{eq:apprhoij}) \cite{Kosowsky:1994cy}, (Fig.(\ref{fig2})). More precisely, the evolution of polarization terms of CMB is indicated by Eq. (\ref{eq:appcollision}). All coefficients appear in the mentioned equations are defined below:
   \begin{eqnarray}\label{eq:appF}
   F_{QT}^{k}&\equiv&2[|\hat{p}.\epsilon_{1}(\vec{k})|^{2}-|\hat{p}.\epsilon_{2}(\vec{k})|^{2}]\nonumber\\
   F_{QT}^{p}&\equiv&4[|\epsilon_{1}.\epsilon_{2}|^{^{2}}-|\epsilon_{2}.\epsilon_{2}|^{2}+|\epsilon_{1}(\vec{k}).\epsilon_{2}|^{2}-|\epsilon_{1}(\vec{k}).\epsilon_{1}(\vec{p})|^{2}]\nonumber\\
   F_{QQ}^{p}&\equiv&4\{  [ \epsilon_{1}(\vec{k})\epsilon_{1}(\vec{p})\epsilon_{1}(\vec{k})\epsilon_{2}(\vec{p})+|\epsilon_{1}(\vec{k}).\epsilon_{1}(\vec{p})|^{2}-|\epsilon_{1}(\vec{k}).\epsilon_{2}(\vec{p})|^{2}]\nonumber\\&-&[\epsilon_{2}(\vec{k}).\epsilon_{2}(\vec{p}) \epsilon_{2}(\vec{k})\epsilon_{1}(\vec{p})+|\epsilon_{2}(\vec{k}).\epsilon_{1}(\vec{k})|^{2}-|\epsilon_{2}(\vec{k}).\epsilon_{2}(\vec{p})|^{2}]\},\nonumber\\
   F_{UT}^{k}&\equiv&4[\hat{p}.\epsilon_{1}(\vec{k})\hat{p}\epsilon_{2}(\vec{k})+p.\epsilon_{1}p.\epsilon_{2}] \nonumber\\ F_{UQ}^{p}&\equiv&4[p.\epsilon_{1}(k)p.\epsilon_{2}(k)-p.\epsilon_{1}(k)p.\epsilon_{2}] \nonumber\\
   F_{UU}^{p}&\equiv&4[p.\epsilon_{1}(k)p.\epsilon_{1}(k)-p.\epsilon_{2}(k)p.\epsilon_{2}] \nonumber\\
   F_{UT}^{p}&\equiv& 4\{[\epsilon_{1}.\epsilon_{1}\epsilon_{2}(\vec{k}).\epsilon_{1}(\vec{p})+\epsilon_{2}.\epsilon_{2}\epsilon_{1}(\vec{k}).\epsilon_{2}(\vec{p})]+[\epsilon_{1}.\epsilon_{1}\epsilon_{2}(\vec{k}).\epsilon_{1}(\vec{p})+\epsilon_{2}.\epsilon_{2}\epsilon_{1}(\vec{k}).\epsilon_{2}(\vec{p})]\}\nonumber\\
   F_{UQ}^{p}&=&4\{ [\epsilon_{1}.\epsilon_{1}\epsilon_{2}.\epsilon_{2}+\epsilon_{1}(\vec{k}).\epsilon_{2}(\vec{p})\epsilon_{2}(\vec{k}).\epsilon_{1}(\vec{p})+\epsilon_{1}(\vec{k}).\epsilon_{1}(\vec{p})\epsilon_{2}(\vec{k}).\epsilon_{1}(\vec{p})-\epsilon_{1}(\vec{k}).\epsilon_{2}(\vec{p})\epsilon_{2}(\vec{k}).\epsilon_{2}(\vec{p})]\nonumber\\&+&[\epsilon_{1}.\epsilon_{1}\epsilon_{2}.\epsilon_{2}+\epsilon_{1}(\vec{k}).\epsilon_{2}(\vec{p})\epsilon_{2}(\vec{k}).\epsilon_{1}(\vec{p})+\epsilon_{1}(\vec{k}).\epsilon_{1}(\vec{p})\epsilon_{2}(\vec{k}).\epsilon_{1}(\vec{p})-\epsilon_{1}(\vec{k}).\epsilon_{2}(\vec{p})\epsilon_{2}(\vec{k}).\epsilon_{2}(\vec{p})]\}
   \end{eqnarray}
   Matter perturbations in the Fourier modes are characterized by the wave vector $\vec{K}$ and the coordinate system $\hat{K} \|\hat{z} $, and their amplitudes depend on the angle between the photon direction and the wave vector $\mu=\hat{n}.\hat{K}$. The Eq. (\ref{eq:appcollision}) is in a standard scenario written without considering dipole asymmetry.
   To take into account the contribution of dipole asymmetry (Eq. (\ref{eq:tamplate})), the $\Delta_{T} (\vec{p},K)$ and $\Delta_{T} (\vec{k},K)$ should be replaced by $ \tilde{\Delta}_{T} (\vec{p},K)= \Delta_T(\vec{p},K)(1+A_T\hat{P}\cdot\hat{p}) $ and $ \tilde{\Delta}_{T} (\vec{k},K)=\Delta_T(\vec{k},K)(1+A_T\hat{P}.\hat{k})$,  respectively in the Boltzmann equations (Eq. (\ref{eq:appcollision})). Therefore, the evolution equation for Stokes parameters, $Q^{(S)}$ and $U^{(S)}$ are modified in the presence of dipole asymmetric scalar perturbations as:
    \begin{eqnarray}
    \frac{d}{d\tau}(\tilde{Q}^{(S)}+ i\tilde{U}^{(S)}) +iK\mu (\tilde{Q}^{(S)}\pm i\tilde{U}^{(S)}) &=& \dot\kappa[
    -(\tilde{Q}^{(S)}\pm i\tilde{U}^{(S)}) -{1\over 2} [1-P_2(\mu)] \Pi+\,\Pi^{+ (S)}(K,\tau)]\nonumber\\
    \frac{d}{d\tau}(\tilde{Q}^{(S)}- i\tilde{U}^{(S)}) +iK\mu (\tilde{Q}^{(S)}\pm i\tilde{U}^{(S)}) &=& \dot\kappa[
    -(\tilde{Q}^{(S)}\pm i\tilde{U}^{(S)}) -{1\over 2} [1-P_2(\mu)] \Pi+\,\Pi^{- (S)}(K,\tau)],\nonumber\\
    \label{Boltzmann1A}
    \end{eqnarray}
    where the last terms in the right-hand side of the above equations in the symmetric scalar perturbations vanished and therefore they are produced due to the asymmetric part. Mentioned terms are as follows:
    \begin{equation}\label{Apm}
    \Pi^{\pm (S)}(K,\tau)=A_T\int \frac{d\Omega_p}{4\pi}\,(\hat{p}\cdot\hat{P})\,\Delta_T(\vec{p},K,\tau)\,(F_{QT}^p\pm i\,F_{UT}^p),
    \end{equation}
    where $\hat{p}\cdot\hat{P}$ has below general form
    \begin{equation}\label{pP1}
    \hat{p}\cdot\hat{P}=\frac{4\pi}{3}\sum\,Y^*_{1,m}(\hat P)Y_{1,m}(\hat p).
    \end{equation}
    To go further, plugging the Eq. (\ref{pP1}) into Eq. (\ref{Apm}), we obtain
    \begin{equation}\label{Apm1}
    \Pi^{\pm (S)}(K,\tau)=A_T\sum_{m'=-3}^3\,\Delta_{T,3m'}\,F^\pm_{3m'} (\hat{n},\hat{P}),
    \end{equation}
    Therefore, the necessary functions to clarify the evolution of Stokes parameters which are given by Eq. (\ref{Boltzmann1A}) are:
   \begin{eqnarray}\label{eq:appf3}
   &&F^\pm_{30} = A_{1}(\hat{P}) Y_{2\,-1}(\hat n)+A_{2}(\hat{P}) Y_{2\,0}(\hat n) +A_{3}(\hat{P}) Y_{2\,1}(\hat n)\nonumber\\
   &&F^\pm_{31} = B_{1}(\hat{P}) Y_{2\,-2}(\hat n)+B_{2}(\hat{P}) Y_{2\,-1}(\hat n)+B_{3}(\hat{P})Y_{2\,0}(\hat n)\pm B_{4}(\hat{P}) Y_{3\,-2}(\hat n)+B_{5}(\hat{P}) Y_{4\,-2}(\hat n)\nonumber\\
   &&F^\pm_{3-1} = C_{1}(\hat{P}) Y_{2\,0}(\hat n)+C_{2}(\hat{P}) Y_{2\,1}(\hat n)+C_{3}(\hat{P}) Y_{2\,2}(\hat n)\pm C_{4}(\hat{P}) Y_{3\,2}(\hat n)+C_{5}(\hat{P}) Y_{4\,2}(\hat n)\nonumber\\
   &&F^\pm_{3-2}= D_{1}(\hat{P}) Y_{2\,1}(\hat n)+D_{2}(\hat{P}) Y_{2\,2}(\hat n)\mp D_{3}(\hat{P}) Y_{3\,2}(\hat n)+D_{4}(\hat{P}) Y_{4\,2}(\hat n)\nonumber\\
   &&F^\pm_{32}= E_{1}(\hat{P}) Y_{2\,-2}(\hat n)+E_{2}(\hat{P}) Y_{2\,-1}(\hat n)\pm E_{3}(\hat{P}) Y_{3\,-2}(\hat n)+E_{4}(\hat{P}) Y_{4\,-2}(\hat n)\nonumber\\
   &&F^\pm_{3-3} =F_{1}(\hat{P}) Y_{2\,2}(\hat n)\mp F_{2}(\hat{P})Y_{3\,2}(\hat n)+F_{3}(\hat{P}) Y_{4\,2}(\hat n)\nonumber\\
   &&F^\pm_{33} =G_{1}(\hat{P})Y_{2\,-2}(\hat n)\mp G_{2}(\hat{P})Y_{3\,-2}(\hat n)+G_{3}(\hat{P})Y_{4\,-2}(\hat n)
   \end{eqnarray}
   The coefficients $ A_{1}, A_{2} ,... G_{3}$ in the above equations read as:
   \begin{eqnarray}
   A_{1}(\hat{P})&=&- \dfrac{8\pi}{5}\sqrt{\dfrac{2}{105}}\sqrt{\dfrac{2\pi}{3}} Y_{1\,-1}(\hat P)~~~,~~~
   A_{2}(\hat{P})=  \dfrac{16\pi}{5\sqrt{35}}\sqrt{\dfrac{\pi}{3}} Y_{1\,0}(\hat P)~~~~,~~~~
   A_{3}(\hat{P})= - \dfrac{8\pi}{5}\sqrt{\dfrac{2}{105}}\sqrt{\dfrac{2\pi}{3}} Y_{1\,1}(\hat P)\nonumber\\ B_{1}(\hat{P})&=& \dfrac{-2A_T}{5\sqrt{70}}\sqrt{\dfrac{2\pi}{3}} Y_{1\,-1}(\hat P)~~~~~~~~,~~~~~~~
   B_{2}(\hat{P})=  \dfrac{-8}{15}\sqrt{\dfrac{2}{35}} \sqrt{\dfrac{2\pi}{3}} Y_{1\,0}(\hat P)~~,~
   B_{3}(\hat{P})= \dfrac{4}{5\sqrt{105}}\sqrt{\dfrac{2\pi}{3}} Y_{1\,1}(\hat P)\nonumber\\
   B_{4}(\hat{P})&=& \dfrac{2}{30\sqrt{10}}\sqrt{\dfrac{2\pi}{3}} Y_{1\,-1}(\hat P)~~~~~~,~~~~~~~ B_{5}(\hat{P})=- \dfrac{2}{5\sqrt{210}}\sqrt{\dfrac{2\pi}{3}} Y_{1\,-1}(\hat P)~~,~
   C_{1}(\hat{P})= \dfrac{4}{5\sqrt{105}}\sqrt{\dfrac{2\pi}{3}} Y_{1\,-1}(\hat P)
   \nonumber\\C_{2}(\hat{P})&=&  \dfrac{-8}{15}\sqrt{\dfrac{2}{35}} \sqrt{\dfrac{2\pi}{3}} Y_{1\,0}(\hat P)~~~,~~~~~~~
   C_{3}(\hat{P})= -\dfrac{2}{5\sqrt{70}}\sqrt{\dfrac{2\pi}{3}} Y_{1\,1}(\hat P)~~~,~~~
   C_{4}(\hat{P})=-\dfrac{2}{30\sqrt{10}}\sqrt{\dfrac{2\pi}{3}} Y_{1\,1}(\hat P)\nonumber\\ C_{5}(\hat{P})&=&- \dfrac{2}{5\sqrt{210}}\sqrt{\dfrac{2\pi}{3}} Y_{1\,1}(\hat P)~~~~~,~~~~~~D_{1}(\hat{P})=-\dfrac{4}{15\sqrt{7}}\sqrt{\dfrac{2\pi}{3}} Y_{1\,-1}(\hat P)~~~~,~~
   D_{2}(\hat{P})=-\dfrac{2}{5\sqrt{7}}\sqrt{\dfrac{2\pi}{3}} Y_{1\,0}(\hat P)\nonumber\\ D_{3}(\hat{P})&=&\dfrac{2}{30}\sqrt{\dfrac{2\pi}{3}} Y_{1\,0}(\hat P)~~~~~~~~~~~~,~~~~~D_{4}(\hat{P})=-\dfrac{2}{5\sqrt{21}}\sqrt{\dfrac{2\pi}{3}} Y_{1\,0}(\hat P) ~~~~~,~~~~
   E_{1}(\hat{P})=-\dfrac{2}{5\sqrt{7}}\sqrt{\dfrac{2\pi}{3}} Y_{1\,0}(\hat P)\nonumber\\ E_{2}(\hat{P})&=&- \dfrac{4}{15\sqrt{7}}\sqrt{\dfrac{2\pi}{3}} Y_{1\,1}(\hat P)~~~~~~,~~~~~E_{3}(\hat{P})=\dfrac{2}{30}\sqrt{\dfrac{2\pi}{3}} Y_{1\,0}(\hat P)~~~~~~~~,~~~~~~~~
   E_{4}(\hat{P})=-\dfrac{2}{5\sqrt{21}}\sqrt{\dfrac{2\pi}{3}} Y_{1\,0}(\hat P)\nonumber\\ F_{1}(\hat{P})&=&- \dfrac{2}{5}\sqrt{\dfrac{3}{14}}\sqrt{\dfrac{2\pi}{3}} Y_{1\,-1}(\hat P)~~~,~~~F_{2}(\hat{P})=\dfrac{2}{10\sqrt{6}}\sqrt{\dfrac{2\pi}{3}} Y_{1\,-1}(\hat P)~~~,~~~~~~~~ F_{3}(\hat{P})=-\dfrac{2}{5\sqrt{14}}\sqrt{\dfrac{2\pi}{3}} Y_{1\,-1}(\hat P)\nonumber\\G_{1}(\hat{P})&=&- \dfrac{2}{5}\sqrt{\dfrac{3}{14}}\sqrt{\dfrac{2\pi}{3}} Y_{1\,1}(\hat P)~~~,~~~~~G_{2}(\hat{P})=-\dfrac{2}{10\sqrt{6}}\sqrt{\dfrac{2\pi}{3}} Y_{1\,1}(\hat P)~~~~~,~~~~~~ G_{3}(\hat{P})=-\dfrac{2}{5\sqrt{14}}\sqrt{\dfrac{2\pi}{3}} Y_{1\,1}(\hat{P})\nonumber\\
   \end{eqnarray}

   \subsection*{Appendix B}\label{appB}
   To compute  the $B$-mode power spectrum and to achieve the Eq. (\ref{BModeM}), we need to clarify following term:
   \begin{eqnarray}\label{62}
   &&   \bar{\eth}^{2}\Pi^{+(\rm S)}(K,\tau) e^{ix\mu}-\eth^{2}\Pi^{- (\rm S)} (K,\tau)e^{ix\mu}= A_T \Delta_{30}[\bar{\eth}^{2}F^{+}_{30}  e^{ix\mu}-\eth^{2}F^{-}_{30} e^{ix\mu}] + A_T \Delta_{31}[\bar{\eth}^{2} F^{+}_{31} e^{ix\mu}-\eth^{2} F^{-}_{31} e^{ix\mu}]\nonumber\\ && +  A_T \Delta_{3-1}[\bar{\eth}^{2} F^{+}_{3-1}e^{ix\mu}-\eth^{2}F^{-}_{3-1} e^{ix\mu}]+ A_T \Delta_{3-2}[\bar{\eth}^{2} F^{+}_{3-2}e^{ix\mu}-\eth^{2}F^{-}_{3-2}e^{ix\mu}] \nonumber\\ &&+ A_T \Delta_{32}[\bar{\eth}^{2} F^{+}_{32}e^{ix\mu}-\eth^{2}F^{-}_{32}e^{ix\mu}] + A_T \Delta_{3-3}[\bar{\eth}^{2} F^{+}_{3-3}e^{ix\mu}-\eth^{2}F^{-}_{3-3}e^{ix\mu}] \nonumber\\ &&+ A_T \Delta_{33}[\bar{\eth}^{2} F^{+}_{33}e^{ix\mu}-\eth^{2}F^{-}_{33}e^{ix\mu}] \nonumber\\ &=& A_T \Delta_{30}[\bar{\eth}^{2}F^{+}_{30} -\eth^{2}F^{-}_{30} ] e^{ix\mu}  + A_T\Delta_{30} [F^{+}_{30}~\bar{\eth}^{2} e^{ix\mu}-F^{-}_{30}~\eth^{2}e^{ix\mu}]\nonumber\\ &&+A_T \Delta_{31}[\bar{\eth}^{2} F^{+}_{31} -\eth^{2} F^{-}_{31} ] e^{ix\mu}  + A_T \Delta_{31}[F^{+}_{31} \bar{\eth}^{2} e^{ix\mu}  - F^{-}_{31} \eth^{2} e^{ix\mu} ]\nonumber\\ &&+ A_T \Delta_{3-1}[\bar{\eth}^{2} F^{+}_{3-1}-\eth^{2}F^{-}_{3-1} ]e^{ix\mu}  +  A_T \Delta_{3-1}[F^{+}_{3-1}\bar{\eth}^{2}e^{ix\mu} -F^{-}_{3-1} \eth^{2} e^{ix\mu} ] \nonumber\\ && + A_T \Delta_{3-2}[\bar{\eth}^{2} F^{+}_{3-2}-\eth^{2}F^{-}_{3-2} ]e^{ix\mu}  + A_T \Delta_{3-2}[F^{+}_{3-2}\bar{\eth}^{2}e^{ix\mu} -F^{-}_{3-2} \eth^{2} e^{ix\mu} ]  \nonumber\\ &&  + A_T \Delta_{32}[\bar{\eth}^{2} F^{+}_{32}-\eth^{2}F^{-}_{32} ]e^{ix\mu}  +  A_T \Delta_{32}[F^{+}_{32}\bar{\eth}^{2}e^{ix\mu} -F^{-}_{32} \eth^{2} e^{ix\mu} ]  \nonumber\\ && + A_T \Delta_{3-2}[\bar{\eth}^{2} F^{+}_{33}-\eth^{2}F^{-}_{3-2} ]e^{ix\mu}  +  A_T \Delta_{33}[F^{+}_{33}\bar{\eth}^{2}e^{ix\mu} -F^{-}_{33} \eth^{2} e^{ix\mu} ] \nonumber\\ && + A_T \Delta_{3-3}[\bar{\eth}^{2} F^{+}_{3-3}-\eth^{2}F^{-}_{3-3} ]e^{ix\mu}  +  A_T \Delta_{3-3}[F^{+}_{3-3}\bar{\eth}^{2}e^{ix\mu} -F^{-}_{3-3} \eth^{2} e^{ix\mu} ]\nonumber\\
   \end{eqnarray}
   Because in the $ \vec{K} \mid\mid z$ coordinate frame, $ U $ and $Q$ are only a function of $\mu$ so
   $ \bar{\eth}^{2}=\eth^{2}$.
   By using  the identity $\exp( i \vec{k}_1 \cdot \vec{x} )
   =  \sum_\ell (-i)^\ell \sqrt{4\pi(2\ell+1)} j_\ell(k_1r)
   Y_\ell^0(\hat{n})$, and
   \begin{eqnarray}
    _{s}Y_{\ell m}&=&\left[\dfrac{(\ell-s)!}{(\ell+s)!}\right]^{\dfrac{1}{2}}~\eth^{s} Y_{\ell m} ~~~~~,~~~~~ (0\leq s \leq \ell) \nonumber\\ _{s}Y_{\ell m}&=&\left[\dfrac{(\ell+s)!}{(\ell-s)!}\right]^{\dfrac{1}{2}}~ (-1)^{s}\bar{\eth}^{-s} Y_{\ell m} ~~~~,~~~~ (-\ell \leq s \leq 0)
   \end{eqnarray}
   according to the equation of $(A3)$ of reference \cite{Zaldarriaga:1996xe}, the Eq. (\ref{62})  becomes:
   \begin{eqnarray}\label{eq:bterm}
    &&\bar{\eth}^{2}\Pi^{+(\rm S)}(K,\tau) e^{ix\mu}-\eth^{2}\Pi^{- (\rm S)} (K,\tau)e^{ix\mu}=  (5!)^{1/2}A_T(B_{4}\Delta_{31}-E_{3}\Delta_{32}-G_{2}\Delta_{33}) \big[ _{-2}Y_{32}+~_{2}Y_{32}\big] \nonumber\\&&+
    (5!)^{1/2}A_T(C_{4}\Delta_{3-1}-D_{3}\Delta_{3-2}-F_{2}\Delta_{3-3}) \big[ _{-2}Y_{3-2}+~_{2}Y_{3-2}\big] \nonumber\\ &&+ A_T(B_{4}\Delta_{31}+E_{3}\Delta_{32}-G_{2}\Delta_{33})Y_{3-2}(\hat{n})\big[\bar{\eth}^{2}e^{ix\mu} + \eth^{2} e^{ix\mu}\big]  \nonumber\\ &&+ A_T(C_{4}\Delta_{3-1}-D_{3}\Delta_{3-2}-F_{2}\Delta_{3-3})Y_{32}(\hat{{n}})\big[\bar{\eth}^{2}e^{ix\mu} + \eth^{2} e^{ix\mu}\big] \nonumber\\
    &=&H_{5}(\hat{P})  \big[ _{-2}Y_{3-2}+~_{2}Y_{3-2}\big] + H_{6}(\hat{P})
    \big[ _{-2}Y_{32}+~_{2}Y_{32}\big]+H_{9}(\hat{P})\big[\bar{\eth}^{2}e^{ix\mu} + \eth^{2} e^{ix\mu}\big]
   \end{eqnarray}
   where
   \begin{eqnarray}
    H_{5}(\hat{P}) &=&(5!)^{1/2}A_T(B_{4}\Delta_{31}-E_{3}\Delta_{32}-G_{2}\Delta_{33}) \nonumber\\
    H_{6}(\hat{P}) &=& (5!)^{1/2}A_T(C_{4}\Delta_{3-1}-D_{3}\Delta_{3-2}-F_{2}\Delta_{3-3}) \nonumber\\
    H_{9}(\hat{P}) &=& A_T (B_{4}\Delta_{31}+E_{3}\Delta_{32}-G_{2}\Delta_{33})Y_{3-2}(\hat{n})+ A_T(C_{4}\Delta_{3-1}-D_{3}\Delta_{3-2}-F_{2}\Delta_{3-3})Y_{32}(\hat{{n}})
   \end{eqnarray}
   we also implement following relations:
   \begin{eqnarray}
    &&[\bar{\eth}^{2} +\eth^{2} ]e^{ix\mu}=\sum_{\ell=2} (-i)^{2} \sqrt{4\pi (2\ell+1)\dfrac{(\ell+2)!}{(\ell-2)!}}j_{\ell}(x)\big[ _{-2}Y_{\ell0}+~_{2}Y_{\ell0}\big] \nonumber\\
   \end{eqnarray}
   \begin{eqnarray}
    _{2}Y_{3\pm2}+~_{-2}Y_{3\pm2}&=& \dfrac{1}{4}\sqrt{\dfrac{7}{\pi}}  e^{\pm2i\varphi}(2\cos^{2}\theta-1)\cos \theta
   \end{eqnarray}
   finally, Eq. (\ref{eq:bterm}) reads as:
   \begin{eqnarray}
    \bar{\eth}^{2}\Pi^{+(\rm S)}(K,\tau) e^{ix\mu}-\eth^{2}\Pi^{- (\rm S)} (K,\tau)e^{ix\mu}&=&\dfrac{1}{4}\sqrt{\dfrac{7}{\pi}} e^{-2i\phi}(2\mu^{2}-1)\mu H_{5}(\hat{P})   + \dfrac{\sqrt{7}}{4 \sqrt{\pi}} e^{2i\phi}(2\mu^{2}-1)\mu H_{6}(\hat{P})
    \nonumber\\&&
    \nonumber\\&&+H_{9}(\hat{P})\big[\bar{\eth}^{2}e^{ix\mu} + \eth^{2} e^{ix\mu}\big]
   \end{eqnarray}
   Therefore the asymmetric $B$-mode power spectrum is written by:
   \begin{eqnarray} \label{CB}
    C_{BB,\ell}^{(\rm S)}\Big|_{\rm Asymmetry}&=& \dfrac{(4\pi)}{2\ell+1} \frac{(\ell-2)!}{(\ell+2)!}\int k^{2}dk P_{\varphi}(k) \sum_m \Big|\int d\Omega Y_{\ell m}^{\ast}(\hat{n}) \int_{0}^{\tau_{0}}d\tau \, g(\tau)\, [\hat{\zeta} (\hat{P}, x) - i \hat{\rho}(\hat{P}, x)]   e^{ix\mu}\big|^2 \nonumber\\
   \end{eqnarray}
   where $\hat{\zeta} (\hat{P}, x)  $ and $ \hat{\rho} (\hat{P}, x) $ are:
   \begin{eqnarray}
    \hat{\zeta} (\hat{P}, x)&=& \sum_{\ell=2} (-i)^{2} \sqrt{4\pi (2\ell+1)\dfrac{(\ell+2)!}{(\ell-2)!}}j_{\ell}(x)\big[ _{-2}Y_{\ell0}+~_{2}Y_{\ell0}\big] \nonumber\\
    \hat{\rho} (\hat{P}, x)&=& -\left[\dfrac{\sqrt{7}}{4 \sqrt{\pi}} H_{5}(\hat{P})  (2\partial_{x}^{3}+\partial_{x})  \right]e^{-2i\varphi}-\left[\dfrac{\sqrt{7}}{4 \sqrt{\pi}} H_{6}(\hat{P})  (2\partial_{x}^{3}+\partial_{x})  \right]e^{2i\varphi}
   \end{eqnarray}
   it turns out that the $ \hat{\zeta} (\hat{P}, x)$ does not contain the asymmetric $B$-mode power spectrum. Then Eq. (\ref{CB}) is given by:
   \begin{eqnarray}
    C_{BB,\ell}^{(\rm S)}\Big|_{\rm Asymmetry}&=& (4\pi)^{2} \frac{(\ell+2)!}{(\ell-2)!} \int K^{2}dK P_{\varphi}(K) \left[  \int_{0}^{\tau_{0}}d\tau \, g(\tau)\, \hat{\beta}(\hat{P}, x) \right]^2 \nonumber\\
   \end{eqnarray}
   in which  $ \hat{\beta}(\hat{P}, x)\equiv\dfrac{1}{4} \sqrt{\dfrac{7}{\pi}} \left[H_{5}(\hat{P})+H_{6}(\hat{P})\right] [\ell (1-\ell)(1+\ell)(2+\ell)] \dfrac{(\ell-2)j_{\ell}(x)-xj_{\ell+1}(x)}{x^{3}} $.
   By implementing $j_{\ell+1}(x)=\dfrac{\ell}{x}j_{\ell}(x)-j_{\ell}^{\prime}(x) $, finally,  we achieve:
   \begin{eqnarray}
    C_{BB,\ell}^{(\rm S)}\Big|_{\rm Asymmetry}&=& (4\pi)^{2} \frac{(\ell+2)!}{(\ell-2)!} \int k^{2}dk P_{\varphi}(k) \left[\int_{0}^{\tau_{0}}d\tau \, g(\tau)\, A(K,\tau)\Big|_{\rm Asymmetry} \, \dfrac{(\ell-2)j_{\ell}(x)-xj_{\ell+1}(x)}{x^{3}}\right]^2,\quad\quad \label{Bmode}
   \end{eqnarray}
   where
   \begin{eqnarray}\label{32}
    A(K,\tau)\Big|_{\rm Asymmetry} = \dfrac{1}{4}\sqrt{\dfrac{7}{\pi}}A_T \left[  B_{4}\Delta_{31,T}-E_{3}\Delta_{32,T}-G_{2}\Delta_{33,T}+C_{4}\Delta_{3-1,T}- D_{3}\Delta_{3-2,T}-F_{2}\Delta_{3-3,T}\right].\nonumber\\
   \end{eqnarray}
   We assume  $\Delta_{31,T}=\Delta_{32,T}=\Delta_{33,T}=\Delta_{3-1,T}=\Delta_{3-2,T}=\Delta_{3-3,T}=\Delta_{3,T}$, therefore by considering  $\hat{P}=(218 , -19)\pm29$ we have:
   \begin{eqnarray}
    A(K,\tau)\Big|_{\rm Asymmetry} &=& A_T \Delta_{3,T}\dfrac{16\pi}{4}\sqrt{\dfrac{14}{3}}\left[0.060\left( Y_{1\,1}(\hat P)- Y_{1\,-1}(\hat P)\right)-\dfrac{2}{15} Y_{1\,0}(\hat P)\right]\nonumber\\
    &=&  A_T \Delta_{3,T}(4 \pi)\sqrt{\dfrac{14}{3}} (-0.0239+ 0.0517)
    \nonumber\\&=&  0.753 A_T\Delta_{3,T}
   \end{eqnarray}
     We will consider this result to estimate the order of magnitude of $E$- and $B$-modes with the assumption of dipole asymmetry.

\end{document}